\documentclass[aps,prd,twocolumn,showpacs,preprintnumbers,amsmath,amssymb,floatfix,nofootinbib]{revtex4}

\usepackage{txfonts}
\usepackage{graphicx}
\usepackage{dcolumn}
\usepackage{bm}
\usepackage{amssymb}
\usepackage{latexsym}

\def\be{\begin{equation}}
\def\ee{\end{equation}}
\def\bea{\begin{eqnarray}}
\def\eea{\end{eqnarray}}

\begin{document}

\title{A New Equation of State for Dark Energy Model}
\author{Lei Feng$^{1,3}$}
\author{Tan Lu$^{2,3}$}
\affiliation{$^1$Department of Physics, Nanjing University, Nanjing
210093, China\\$^2$Purple Mountain Observatory, Chinese Academy of
Sciences, Nanjing 210008, China\\$^3$Joint Center for Particle,
Nuclear Physics and Cosmology,
  Nanjing University -- Purple Mountain Observatory,
  Nanjing  210093,
  China}

\begin{abstract}
A new parameterization for the dark energy equation of state(EoS) is
proposed and some of its cosmological consequences are also
investigated. This new parameterization is the modification of
Efstathiou' dark energy EoS parameterization. $w (z)$ is a well
behaved function for $z\gg1$ and
 has same behavior in $z$ at low redshifts with Efstathiou'
parameterization. In this  parameterization there are two free
parameter $w_0$ and $w_a$. We discuss the constraints on this
model's parameters from current observational data. The best fit
values of the cosmological parameters with
 $1\sigma$ confidence-level regions are:
$\Omega_m=0.2735^{+0.0171}_{-0.0163}$,
$w_0=-1.0537^{+0.1432}_{-0.1511}$ and
$w_a=0.2738^{+0.8018}_{-0.8288}$.
\end{abstract}

 \maketitle

\section{Introduction}
In recent years, the discovery of accelerating expansion of the
universe
 is an
amazing development. It was firstly discovered by observing type Ia
supernova (SNe Ia)~\cite{ref1,ref2}, which can be used as standard
candles \cite{ref3,ref4}. The cosmic microwave background (CMB)
measurements from Wilkinson Microwave Anisotropy Probe (WMAP)
\cite{ref5} and the large scale structure survey by Sloan Digital
Sky Survey (SDSS) \cite{ref6,ref7} confirm this accelerating
expansion universe model. There are two kinds of ideas, i.e. the
existence of the dark energy or modifications of the gravity theory,
to explain this concept. The first scheme is most popularly
discussed, and many models have been proposed, such as the
holographic dark energy models \cite{ref9} and the Chaplygin Gas
\cite{ref11} . In addition there are also many modified gravity
models, such as the brane world \cite{ref10} and $f(R)$ \cite{fr}
and so on.

The dark energy equation of state(EoS) which is the ratio of
pressure to energy density, is a prefect quantity to study the
behavior of dark energy. If EoS is a constant $-1$, dark energy is
the $\Lambda$ Cold Dark Matter Model(LCDM), and maybe dark energy is
vacuum energy. Otherwise the dark energy would be dynamical scalar
field, such as the Quintessence \cite{ref8}, the
quintom\cite{quintom}.

There are several way to explore the behavior of dark energy EoS.
The most popular way is to build a functional form for EoS in terms
of some free parameters. Lots of EoS parameterizations have already
been discussed in the literature(such as
\cite{gteg,gHuterer,agstier,Wgeller,Efstathiou,Chgevallier,Lignder,golgiah,otherPg,zhuzonghong,wuyueliang}
and Refs. therein). Another way is picking a simple local basis
representation for $w(z)$ (bins, wavelets), and estimate the
associated coefficients ~\cite{huterer04,hojjati09,daly03}. In
addition, there are also some nonparametric way~\cite{prl,zuixin}.

In this paper, we consider a new parameterization for the dark
energy EoS. In\cite{Efstathiou}, the author developed a new dark
energy EoS parameterization: $w(z) = w_0+w_a\ln(1+z)$. It is obvious
that when $z\rightarrow\infty$, $w(z)$ has poor behavior and becomes
infinite . This dark energy EoS parameterization can only describe
the behavior of dark energy when $z$ is not very large. To avoid
this problem, we consider a new dark energy EoS parameterization,
which is the modification of Efstathiou' parameterization: $w(z) =
w_0+w_a\ln(1+\frac {z}{1+z})$. The value of $w(z)$ is $w_0$ at
present and $w(z)$ becomes to ($w_0+w_a*ln2$) when
$z\rightarrow\infty$. In this model there are three parameters in
all, which is $w_0$, $w_a$ and $\Omega_m$. As shown
in\cite{Ma:2011nc}, this EoS will get to a nonphysical value in the
far future time when redshift $z$ approaches $-1$, namely, $|w(z)|$
will grow rapidly and diverge.

In this paper, we perform a global data fitting analysis on this new
dark energy EoS parametrizations, and present constraints on the
model parameters from the current observational data, including the
 seven-year
WMAP data, Baryon Acoustic Oscillations (BAO) data, Observational
Hubble data and SN Union2 sample. Since dark energy parameters are
tightly correlated to some other cosmological parameters, such as
the matter density parameter $\Omega_m$ and the Hubble constant
$H_0$, it is necessary to consider a global fit procedure in the
investigation of the dynamical dark energy. The paper is organized
as follows: In section II, we review the new dark energy EoS
parameterization. In section III, we describe Current Observational
Data we used. In section IV, we perform the cosmic observation
constraint, the results are also presented. The last section is the
conclusion.

\section{New Parameterization}

Let us start by presenting some of the most investigated EoS
parameterizations(see also \cite{otherPg} for other
parameterizations):
\begin{eqnarray}
\label{OtherParameterizations} w(z) = \; \left\{
\begin{tabular}{l}
$w_0 + w_a z$ 
\, \quad \quad  \quad \quad \mbox{(redshift)} \quad\,  \hspace{0.2cm}\cite{gHuterer,agstier,Wgeller}\\
\\
$w_0+w_az/(1+z)$ \quad \mbox{(scale factor)} \,
\cite{Chgevallier,Lignder}\\
\\
$w_0+w_a\ln(1+z)$ 
\quad  \mbox{(logarithmic)} \quad \cite{Efstathiou}
\end{tabular}
\right. \nonumber
\end{eqnarray}
where $w_0$ is the current value of the EoS, and $w_a$ indicate the
revolution of the dark energy EoS. The value of these parameter is
determined by the observational data. When $w_a = 0$ and $w_0 = -1$,
the dark energy model becomes to LCDM model.

The first parameterization represents a good fit for low redshifts,
but has serious problems to explain high-$z$ observations since it
blows up as $\exp{(3w_az)}$ when $z > 1$ and $w_a > 0$. For example,
it can not explain the estimated ages of high-$z$ objects
\cite{Friaca}. The second one solves this problem, since $w (z)$ is
a well behaved function for $z\gg1$ and recovers the linear behavior
in $z$ at low redshifts. The latter was introduced by
Efstathiou~\cite{Efstathiou}. It was built empirically to adjust
some quintessence models at $z\lesssim4$. When $z$ approachs
infinity, $w(z)$ has poor behavior and becomes infinite and this is
unnatural. Similar to the second model, let us consider the
following EoS parameterization
 \be\label{eqn::eos2} w(z) = w_0+w_a\ln(1+\frac {z}{1+z}). \ee
where the value of $w(z)$ is $w_0$ at present and $w(z)\rightarrow
(w_0+w_a*ln2)$ when $z\rightarrow\infty$.

If there is no interaction between dark energy and other component
of the universe, one can show from the energy conservation law
$[\dot{\rho}=-3\dot{a}(\rho+p)/a ]$ that the dark energy density
evolves as
\begin{equation}
\rho_\textsc{de}(z) = \rho_c (1-\Omega_{m}) \exp \left( 3
\int_{0}^{z} \left[ 1 + w(z') \right] \frac{dz'}{1+z'} \right) ,
\label{eq:rho phi (z)}
\end{equation}
where $\rho_c$ is the critical density, it is defined by the
following equation
\begin{equation}
\rho_c \equiv \frac{3H_0^2}{8 \pi G_N} \, . %
\label{eq:rho0}
\end{equation}
In this new parameterization, it is hard to write out the analysis
formula of this quantity, It is calculated through numerical method.

\begin{figure}
  \includegraphics[width=235pt]{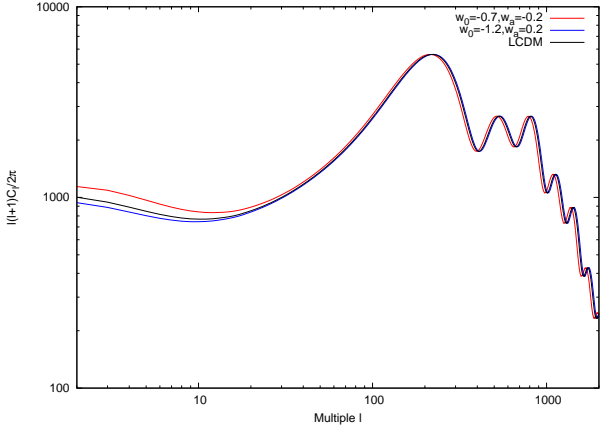}
  \caption{the CMB anisotropy spectrum. Red line: $w_0=-0.7$, $w_a=-0.2$; Green line: $w_0=-1.2$, $w_a=0.2$;
  Black line: $w_0=-1$, $w_a=0$.
 }\label{fig:cls}
\end{figure}
In order to study the evolution of cosmological perturbations, we
use the public Parameterized Post-Friedmann (PPF) package developed
by Wayne Hu(see e.g. \cite{ppf} for detailed discussions on PPF
method) to calculate the CMB anisotropy spectrum. In Figure \ref
{fig:cls} we plot the anisotropy spectrum for different choices of
$w_0$ and $w_a$. We observe some obvious differences with respect to
the LCDM case on larger scales (multipole number $l < 10$). And for
the model $w_0=-0.7~ \mathrm{and}~w_a=-0.2$, there is a slit
deviation on the peaks of anisotropy spectrum.

\section{Current Observational Data}\label{app}
In order to test this new model, we use the most recent
observational data currently available. In this section, we describe
how we use these data.

\subsection{Type Ia Supernovae constraints}

We use the 557 SNe Ia Union2 dataset \cite{union2}. Following
\cite{ref:smallomega,ref:POLARSKI}, one can obtain the corresponding
constraints by fitting the distance modulus $\mu(z)$ as
\begin{equation}
\mu_{th}(z)=5\log_{10}[D_{L}(z)]+\mu_{0}.
\end{equation}
where $\mu_0\equiv 42.38-5\log_{10}h$, and $h$ is the Hubble
constant $H_0$ in units of 100 $\mathrm{km}~
\mathrm{s}^{-1}~\mathrm{Mpc}^{-1}$, In flat universe, the
Hubble-free luminosity distance $D_L=H_0d_L$ is
\begin{equation}
D_L(z)=(1+z)\int_0^z{dz'\over E(z')},
\end{equation}
where $E(z)\equiv H(z)/H_0$.

For the SN Ia dataset, the best fit values of the parameters can be
determined by a likelihood analysis, based on the calculation of \be
\chi^2_{\rm SN}=\sum_i^{557}\frac{[\mu_{\rm th}(z_i)-\mu_{\rm
obs}(z_i)]^2}{\sigma^2(z_i)} \ee

\subsection{Baryon Acoustic Oscillation constraints}

The BAO data come from SDSS DR7 \cite{Percival:2009xn}. The
datapoints we use are
   \be
\frac{r_s(z_d)}{D_V(0.275)}=0.1390\pm0.0037 \ee and \be
\frac{D_V(0.35)}{D_V(0.2)}=1.736\pm0.065 \ee
where the spherical average gives us the following effective distance measure \cite{dvcal},
 \be
D_V(z)=\left [\left(\int_0^z\frac{dx}{H(x)}\right
)^2\frac{z}{H(z)}\right ]^{1/3}  \ee

and $r_s(z_d)$ is the comoving sound horizon at the baryon drag
epoch. Also, $z_d$ can be obtained by using a fitting formula
\cite{ref:Eisenstein}:
\begin{eqnarray}
&&z_d=\frac{1291(\Omega_mh^2)^{0.251}}{1+0.659(\Omega_mh^2)^{0.828}}[1+b_1(\Omega_bh^2)^{b_2}],
\end{eqnarray}
with
\begin{eqnarray}
&&b_1=0.313(\Omega_mh^2)^{-0.419}[1+0.607(\Omega_mh^2)^{0.674}], \\
&&b_2=0.238(\Omega_mh^2)^{0.223}.
\end{eqnarray}

The function $r_s(z)$ is the comoving sound
horizon size
\begin{eqnarray}
r_s(z)=\frac{c}{\sqrt{3}}\int_0^{1/(1+z)}\frac{da}{a^2H(a)\sqrt{1+(3\Omega_b/(4\Omega_\gamma)a)}},
\end{eqnarray}
where $\Omega_\gamma=2.469\times10^{-5}h^{-2}$ for
$T_{CMB}=2.725K$.

So the $\chi^2$ for the BAO
data is given by
\begin{align}
\label{baochi2}
\chi^2_{BAO}=\left(\frac{r_s(z_d)/D_V(z=0.275)-0.1390}{0.0037}\right)^2\nonumber\\
+\left(\frac{D_V(z=0.35)/D_V(z=0.2)-1.736}{0.065}\right)^2.
\end{align}

\subsection{Cosmic Microwave Background constraints}

The CMB shift parameter $R$ is provided by \cite{ref:Bond1997}
\begin{equation}
R(z_{rec})=\frac{\sqrt{\Omega_m
H^2_0}}{\sqrt{|\Omega_k|}}\mathrm{sinn}[\sqrt{|\Omega_k|}\int_0^{z_{rec}}\frac{dz'}{H(z')}],
\end{equation}
where $\mathrm{sinn}(x)$ is $\mathrm{sinh}(x)$ for $\Omega_k > 0$ ,
$x$ for $\Omega_k = 0$, and $\mathrm{sin}(x)$ for $\Omega_k < 0$,
respectively. Here, the redshift $z_{rec}$ (the decoupling epoch of
photons) is obtained by using the fitting function \cite{Hu:1995uz}
\begin{equation}
z_{rec}=1048\left[1+0.00124(\Omega_bh^2)^{-0.738}\right]\left[1+g_1(\Omega_m
h^2)^{g_2}\right],\nonumber
\end{equation}
where
\begin{eqnarray}
g_1&=&0.0783(\Omega_bh^2)^{-0.238}\left(1+ 39.5(\Omega_bh^2)^{0.763}\right)^{-1},\nonumber \\
g_2&=&0.560\left(1+ 21.1(\Omega_bh^2)^{1.81}\right)^{-1}.\nonumber
\end{eqnarray}
In addition, the acoustic scale is related to the distance ratio and
is expressed as
\begin{eqnarray}
&&l_A=\frac{\pi}{r_s(z_{rec})}\frac{c}{\sqrt{|\Omega_k|}}\mathrm{sinn}[\sqrt{|\Omega_k|}\int_0^{z_{rec}}\frac{dz'}{H(z')}].
\end{eqnarray}

Following Ref.~\cite{WMAP7}, the $\chi^2$ for the CMB data is
\begin{equation}
\chi_{CMB}^2=(x^{th}_i-x^{obs}_i)(C^{-1})_{ij}(x^{th}_j-x^{obs}_j),\label{chicmb}
\end{equation}
where $x_i=(l_A, R, z_{rec})$ is a vector and $(C^{-1})_{ij}$ is the
inverse covariance matrix. The seven-year WMAP observations
\cite{WMAP7} give the maximum likelihood values:
$l_A(z_{rec})=302.09$, $R(z_{rec})=1.725$ and $z_{rec}=1091.3$. In
Ref.~\cite{WMAP7}, the inverse covariance matrix is also given as
follows
\begin{equation}
(C^{-1})=\left(
  \begin{array}{ccc}
    2.305 & 29.698 & -1.333 \\
    29.698  & 6825.270 & -113.180 \\
    -1.333  & -113.180  &  3.414 \\
  \end{array}
\right).
\end{equation}

\subsection{Observational Hubble data (OHD)}

The Hubble parameter can be writen as the following form:
\begin{equation}
H(z)=-\frac{1}{1+z}\frac{dz}{dt}.
\end{equation}
So, through measuring $dt/dz$, we can obtain $H(z)$. In
\cite{jimenez_loeb} and \cite{stern, simon}, the author discuss the
possibility of using absolute ages of galaxies to calculate the
value of $dt/dz$. In \cite{simon}, the galaxy spectral data come
from the Gemini Deep Deep Survey \cite{gdds} and archival data
\cite{archivala,archivalb,archivalc,archivald,archivale,archivalf}.
Detailed calculations of $dt/dz$ can be found in \cite{simon}, and
we do not discuss them here. Currently, we have a set of 12 values
of the Hubble parameter versus  redshift in total(see table 2 of
\cite{hzdata}). The data calculated by this way are less sensitive
to systematic errors which is a great advantage\cite{dunlop}.

We can use these data to constraint different kinds of dark energy
models and modified gravity models by minimizing the quantity
 \begin{eqnarray}
\chi^2_{\rm OHD}=\sum_{j=1}^{12}\frac{(H(z_j)-H_{\rm
obs}(z_j))^2}{\sigma_{\rm H,j}^2}.
\end{eqnarray}
This test has already been used to constrain several cosmological
models\cite{L18Yi,Linhui,L19Samushia,L20Wei,L21Zhang,L22Wei,L23Zhang,L24Dantas,L25Zhang,L26Wu,Fenglei,L27Wei,lixin,lixin2}.

\section{Results}

\begin{figure}
  \includegraphics[width=230pt]{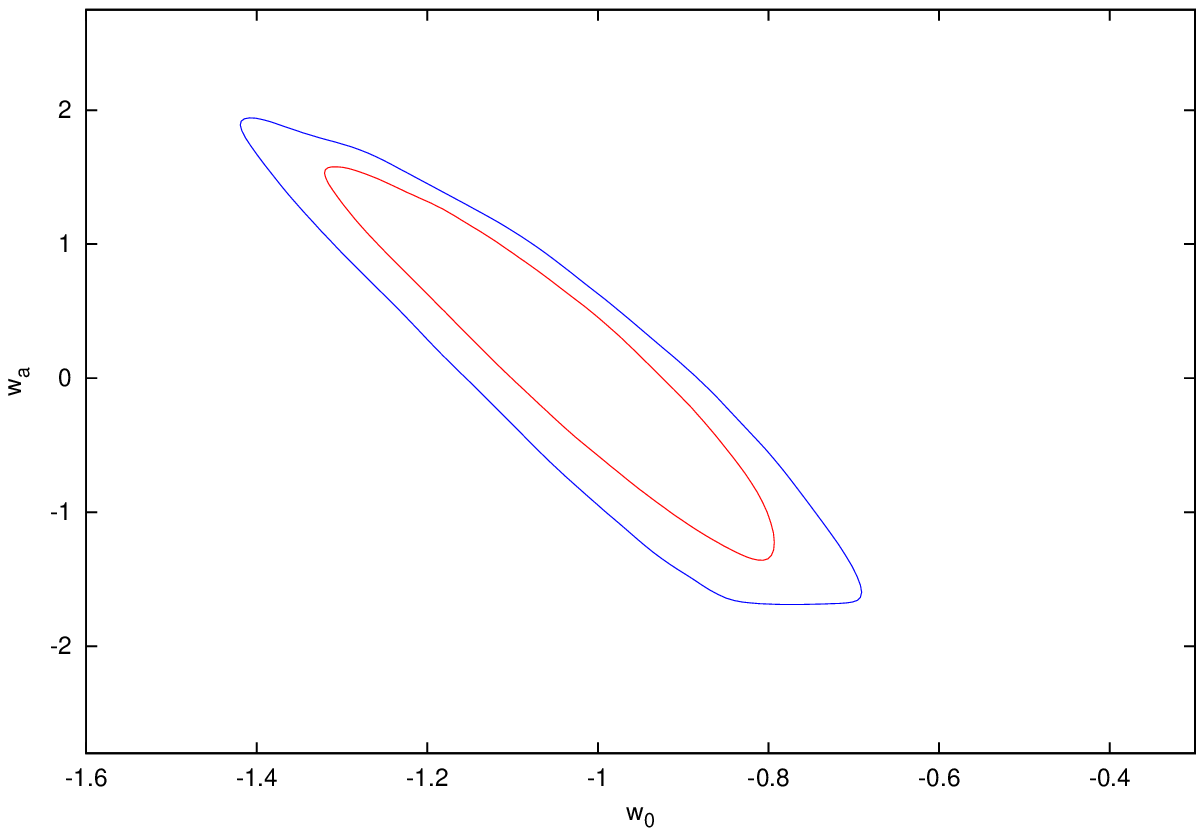}
  \includegraphics[width=235pt]{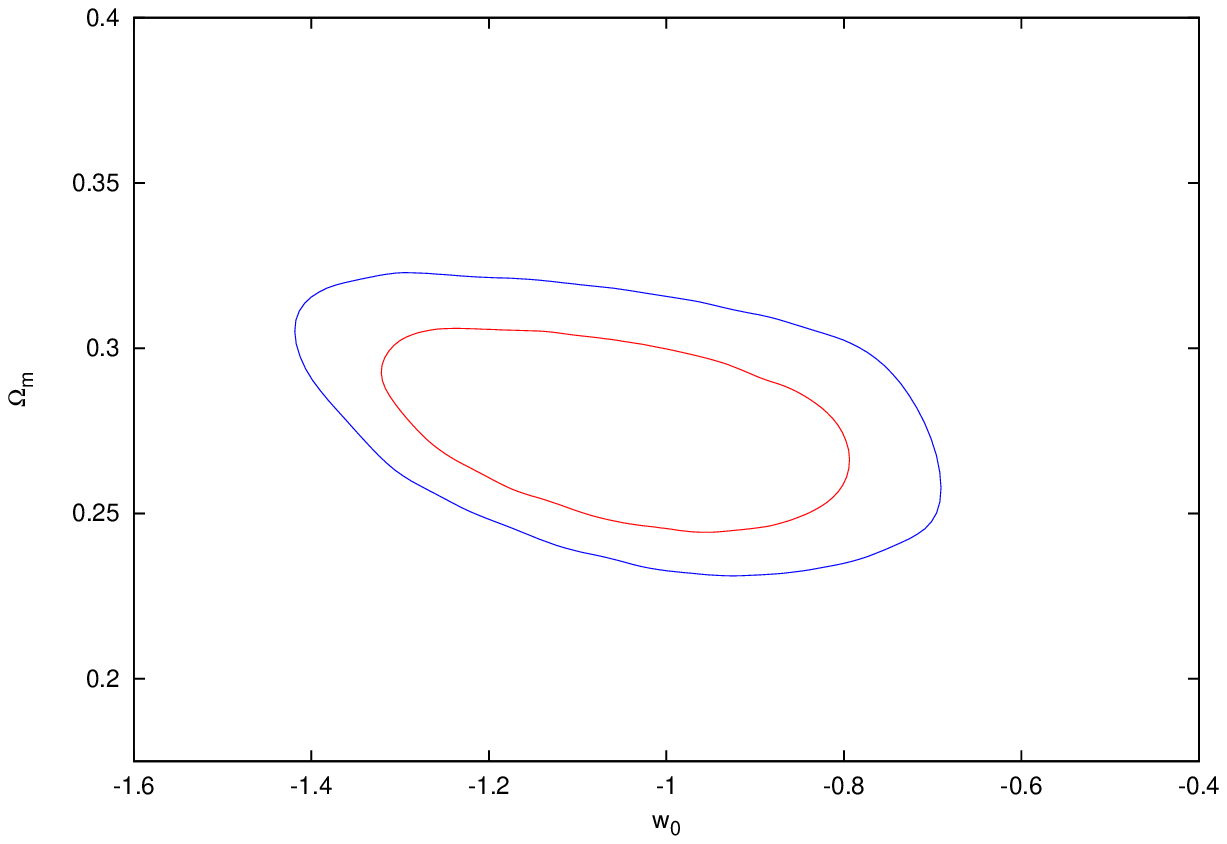}
  \caption{Marginalized probability contours at $1\sigma$ and $2\sigma$ CL in the $\Omega_m-w_0$ and $w_0-w_a$ planes
  for the model considered in this manuscript. The result
is consistent with the LCDM model in the $1\sigma$ CL.
 }\label{fig:flat}
\end{figure}

In our analysis, we perform a global fitting to determine the
cosmological parameters using the MCMC method. In our calculations,
we have taken the total likelihood function $L\propto e^{-\chi^2/2}$
to be the products of the separate likelihoods of SNe Ia, BAO, CMB
and OHD. Then we get $\chi^2$ as
\begin{eqnarray}
\chi^2=\chi^2_{SN}+\chi^2_{BAO}+\chi^2_{CMB}+\chi^2_{OHD},
\end{eqnarray}

The results on the best fit values of the cosmological parameters
with
 $1\sigma$ confidence-level regions are:
$\Omega_m=0.2735^{+0.0171}_{-0.0163}$,
$w_0=-1.0537^{+0.1432}_{-0.1511}$ and
$w_a=0.2738^{+0.8018}_{-0.8288}$. The nuisance parameters $H_0$ used
in the analysis is actually not model parameters with significant
meanings, so we do not list it.

\begin{figure}
  \includegraphics[width=235pt]{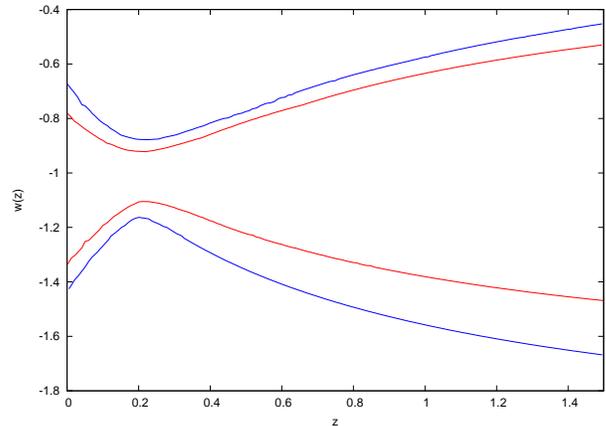}
  \caption{The evolution of $\mathrm{w}(z)$ along with z for the model considered in this manuscript. The result is consistent with the cosmological
  constant in the $1\sigma$ CL.
 }\label{fig:rev}
\end{figure}

In Figures \ref{fig:flat}, we also show the parametric spaces
$w_0-w_a$ and $\Omega_m-w_0$ that arise from the joint analysis
described above. We note that the result is consistent with the
LCDM($w_0=-1$ and $w_a=0$) model in the $1\sigma$ CL. To acquire
more information on the property of dark energy, in Fig.
\ref{fig:rev} we plot the evolution of the EOS $w(z)$ along with
$z$.  The following is some discussion of these findings:
\begin{enumerate}
\item The best-fit results are:
$\Omega_m=0.2735$, $w_0=-1.0537$ and $w_a=0.2738$. Note that here
the results are maximum likelihood values. The value of $w(z)$ is
$-1.0537$ at present and $w(z)$ equals to $-0.8639$ when
$z\rightarrow\infty$. We find that the best-fit dark energy model is
a quintom model \cite{quintom}, whose $w(z)$ crosses the
cosmological constant boundary $w=-1$ during the evolution. And the
redshift is $0.2766$ when $w(z)$ crosses the cosmological constant
boundary $w=-1$.
\item With the
current observational data, the variance of $w_0$ and $w_a$ we get
are still large; the $1\sigma$ constraints on $w_0$ and $w_a$ are
$w_0=-1.0537^{+0.1432}_{-0.1511}$ and
$w_a=0.2738^{+0.8018}_{-0.8288}$. This result implies that though
the dynamical dark energy models are mildly favored, the current
data cannot distinguish different dark energy models decisively.
With the fitting results we obtained, we can reconstruct the
evolution of the EOS of dark energy, $w(z)$ which is shown in
Fig.~\ref{fig:rev}. From the figure, we can directly see that
although the quintom model is more favored, LCDM, however, still
cannot be excluded.
\item From Fig. \ref{fig:rev}, one can see that the allowed value of
$w(z)$ is in the band $\{-1.7,-0.4\}$, which is relatively narrow.

\item The best fit value and $1\sigma$ confidence-level regions of $\Omega_m$ is $0.2735^{+0.0171}_{-0.0163}$,
which is also consistent with the constraint on $\Omega_m$ in the
LCDM model and the CPL model.

\end{enumerate}

\section{Conclusion }
In this paper, we develop a new parameterization which is the
modification of Efstathiou' parameterization. In this new
parameterization, there are three free parameters: $w_0$, $w_a$ and
$\Omega_m$. Then we carry out the global fitting on these model
using the current data: SNe Ia, BAO, CMB and OHD. From the analysis,
the best fit values of the cosmological parameters with
 $1\sigma$ confidence-level regions are:
$\Omega_m=0.2735^{+0.0171}_{-0.0163}$,
$w_0=-1.0537^{+0.1432}_{-0.1511}$ and
$w_a=0.2738^{+0.8018}_{-0.8288}$. From the analysis, we can directly
see that the quintom model is more favored, but this result is also
consistent with the LCDM model in the $1\sigma$ CL.

\section*{Acknowledgments}
Thank Xiaodong Li for helping FL write some figure of this
manuscript.

\end{document}